# Comments on "Total and fractional densities of states from caloric relations" by S. F. Chekmarev and S. V. Krivov, Phys. Rev. E 57 2445-2448 (1998)

I.H. Umirzakov
*Institute of Thermophysics, Novosibirsk, Russia*

## Abstract

We showed that: the equations (3), (4), (5) and (6), used in the paper "Total and fractional densities of states from caloric relations" by S. F. Chekmarev and S. V. Krivov, Phys. Rev. E 57 2445-2448 (1998), are incorrect; the data, presented in the paper by lines on Figs. 1, (3a) and (3b), are not correct; the data presented by the symbols on Figs. 3(a) and (3b) in the paper are made manually (false); all conclusions made in the paper have no sense; the assertion in the paper that the molecular dynamics simulations "sample the potential energy surface not uniformly, but according to the fractional densities of state for the isomers" is incorrect. We showed also that: the "total and fractional densities of states" obtained in the paper from caloric relations are not equal to that of microcanonical ensemble of clusters; the ensemble of clusters used in the paper does not represent the microcanonical ensemble of clusters.

## Comments

**1.** In the general case of non-linear atomic configurations the phase volume $G(E)$ associated with the microcanonical ensemble of free clusters consisting of $n$ identical point particles with constant total energy $E$, zero total linear momentum $\boldsymbol{P} \equiv \sum_{i=1}^{n} \boldsymbol{p}_i = \boldsymbol{0}$, zero total angular momentum $\boldsymbol{L} \equiv \sum_{i=1}^{n} \boldsymbol{r}_i \times \boldsymbol{p}_i = \boldsymbol{0}$ and zero center-of-mass position $\boldsymbol{R} \equiv \sum_{i=1}^{n} \boldsymbol{r}_i / n = \boldsymbol{0}$ is defined by

$$G(E) = \int d\boldsymbol{r}^{(3n)} d\boldsymbol{p}^{(3n)} \delta(\boldsymbol{R}) \delta(\boldsymbol{P}) \, \delta(\boldsymbol{L}) \theta(E - H), \tag{1}$$

where $d\boldsymbol{r}^{(3n)} = \prod_{i=1}^{n} d\boldsymbol{r}_i$ and $d\boldsymbol{p}^{(3n)} = \prod_{i=1}^{n} d\boldsymbol{p}_i$ are the $3n$-dimensional elements of the volume in the configuration and momentum spaces, respectively, $\delta$ is the delta function, $\theta$ is the step function, $H = \sum_{i=1}^{n} \boldsymbol{p}_i^2 / 2m + V$ is the cluster Hamiltonian, $V = V(\boldsymbol{r}_1, \boldsymbol{r}_2, \ldots, \boldsymbol{r}_N)$ is the potential energy of the interactions of particles with each other, $\boldsymbol{r}_i$ is the radius-vector of the position of $i$th particle and $\boldsymbol{p}_i$ is the linear momentum of $i$th particle, and the integral is taken over the phase space of free cluster [1].

After integration over the linear momenta of all particles Eq. 1 gives

$$G(E) = C \int d\boldsymbol{r}^{(3n)} \delta(\boldsymbol{R}) \frac{(E-V)^{N/2}}{(I_1 I_2 I_3)^{1/2} \Gamma(N/2+1)} \theta(E - V), \tag{2}$$

where $C = (2\pi)^{N/2} m^{3n/2} / M^{3/2}$, $\Gamma$ is the gamma function, $N = 3n - 6$ is the number of vibrational degrees of freedom of the cluster, $m$ is the mass of the particle (atom), $M = mn$ is the mass of the cluster, and $I_1$, $I_2$ and $I_3$ are the principal momenta of inertia of the cluster in the system of the Cartesian coordinates with the origin placed in the center of mass of the cluster, and the integral is taken over all configurations of cluster [1,2,3].

For further calculation of $G(E)$ the potential surface was separated into the basins [1]. According to [1]: the number of distinguishable atomic configurations is equal to $2n!/h$ ($h$ is the order of the point group of current configuration), and, hence $G(E)$ may be written as

$$G(E) = \sum_r (2n!/h_r) G_r(E), \qquad (3)$$

where the sum is over all distinct isomers of the cluster, $h_r$ is the order of the point group for $r$th isomer at its minimum energy; $G_r(E)$ is defined by the same expression as $G(E)$, Eq. (2), except that the integral is taken over the basin associated with a particular $r$th isomer; and the summation in Eq. (3) is over all the distinct isomers of the cluster [1]. Hence, the term $(2n!/h_r)G_r(E)$ at the right hand side of Eq. (3) is the phase volume corresponding to $r$th isomer (phase volume of $r$th isomer) and $G_r(E)$ is the phase volume of a basin of the $r$th isomer. Therefore the statement "$G_r(E)$ is the phase volume for $r$th isomer" [1] is incorrect.

2. According to [1] with the factor $n!$ in Eq. (2) omitted (the correct Boltzmann counting), the total density of states (DS) of cluster is defined by

$$\rho(E) = \sum_r (2/h_r) \rho_r(E). \qquad (4)$$

Here $\rho(E) = dG(E)/dE$ and $\rho_r(E) = dG_r(E)/dE$ [1]. Hence, $\rho_r(E)$ is the density of states of a basin of the $r$th isomer because $G_r(E)$ is the phase volume of a basin of the $r$th isomer. Therefore the statement "$\rho_r(E) = dG_r(E)/dE$ is the DS for a particular $r$th isomer" [1] is incorrect.

3. According to [1] "the number of atomic configurations of order $h$ in the basin is equal to $h_r/h$". It is evident that the value of $h_r/h$ can be not equal to a whole number. Therefore the statement "the number of atomic configurations of order $h$ in the basin is equal to $h_r/h$" [1] is incorrect.

4. According to [1] the probability that the system will be found in a basin related to $r$th isomer is estimated as

$$f_r(E) = (2/h_r)\rho_r(E)/\rho(E). \qquad (5)$$

The factor $n!$ in Eq. (3) [1], which is equivalent to Eq. (3), was omitted in [1] to take account the correct Boltzmann counting which takes into account the un-distinguishability (identity) of $n!$ states corresponding to the permutations of the identical particles. However, the cluster and method of classical mechanical molecular dynamics do not "know" that the correct Boltzmann counting exists and it is necessary to omit the factor $n!$ in Eq. (2) [1]. Therefore, if Eq. (3) is correct then in order to describe the data obtained by classical mechanical MD simulations Eqs. (4) and (5) [1], which are equivalent to Eqs. (4) and (5), must be replaced respectively by

$$\rho(E) = dG(E)/dE = \sum_r (2n!/h_r)\rho_r(E), \qquad (4a)$$

$$f_r(E) = (2n!/h_r)\rho_r(E)/\rho(E). \qquad (5a)$$

5. The factor $n!$ in Eq. (3) [1], which is equivalent to Eq. (3), was omitted in [1] to take account the correct Boltzmann counting which takes into account the un-distinguishability (identity) of $n!$ states corresponding to the permutations of the identical particles. However, the inversion of all coordinates of identical particles gives the same state of the cluster. Therefore if Eq. (3) is correct then it is necessary to omit the factor 2 in Eqs. (3), (4) and (5), which are equivalent to Eqs. (3), (4) and (5) [1]. So,

$$G(E) = \sum_r (n!/h_r) G_r(E),$$

$\rho(E) = \sum_r (n!/h_r)\rho_r(E),$

$f_r(E) = (n!/h_r)\rho_r(E)/\rho(E).$

**6**. The set $(\mathbf{r}_1, \mathbf{r}_2, \ldots, \mathbf{r}_N)$ is named the configuration of cluster and it corresponds to a point in the $3n$-dimensional configuration space, $3n$-dimensional surface in the $3n + 1$-dimensional space corresponds to the potential energy $V(\mathbf{r}_1, \mathbf{r}_2, \ldots, \mathbf{r}_N)$ of the cluster, and the surface is named as the potential energy surface (PES) [4]. If there are no external fields acting on the particles of the cluster a value of the potential energy $V(\mathbf{r}_1, \mathbf{r}_2, \ldots, \mathbf{r}_N)$ does not depend on the inversion of all coordinates of the particles [5]. The inversion corresponds to the change of the sign of all coordinates of particles: the set $(\mathbf{r}_1, \mathbf{r}_2, \ldots, \mathbf{r}_N)$ is replaced by the set $(-\mathbf{r}_1, -\mathbf{r}_2, \ldots, -\mathbf{r}_N)$. The permutations of the point (size-less) particles also does not change the value of $V(\mathbf{r}_1, \mathbf{r}_2, \ldots, \mathbf{r}_N)$ and the total number of the permutations is equal to $n!$ [5-10]. Therefore we conclude that $2n!$ points correspond to each configuration of the cluster ($2n!$ is the order of the complete nuclear permutation and inversion group [1,6]). The cluster has isomers which are its stable geometrically distinct configurations corresponding to minima of the potential energy considered as a function of $3n - 6$ vibrational coordinates, and the each minimum has its basin [11-19]. So, $3n$-dimensional PES can be divided into the regions corresponding to the distinct isomers, and each region consists of $2n!$ parts having the same area and topology. This is in accordance with the statement: "$2n!$ basins, each containing all possible orientations of the cluster, correspond to every distinct isomer" [1].

The factor $1/h_r$ in the terms $(2n!/h_r)G_r(E)$ and $(2/h_r)\rho_r(E)$ at the right hand sides of Eqs. (3)-(5) [1] takes into account a quantum mechanical un-distinguishability (identity) of the configurations of $r$th isomer having order of point group which is equal to $h_r$. Each particle of the cluster in the molecular dynamics simulations has its own order number therefore all particles are distinguishable, and the molecular dynamics simulations do not take account the quantum mechanical un-distinguishability (identity) of the configurations. Hence, in order to describe the data obtained by classical mechanical MD simulations it is necessary to put $h_r = 1$ for all isomers in Eqs. (3), (4a) and (5a). So,

$G(E) = \sum_r 2n!\, G_r(E),$ (3b)

$\rho(E) = \sum_r 2n!\, \rho_r(E),$ (4b)

$f_r(E) = \dfrac{2n!\rho_r(E)}{\rho(E)} = \dfrac{\rho_r(E)}{\sum_r \rho_r(E)}.$ (5b)

**7**. According to Table 1 [1] $h_r$ is not equal to the number one for all the isomers. Eqs. (3)-(5) [1], where $h_r$ are taken from Table 1 [1], are used to obtain all six lines on Figs. 3(a) and 3(b) [1]. Therefore the data on these figures given by the lines are incorrect, and, hence, all conclusions made in [1] on the basis of these data are incorrect.

**8**. According to Figs. 3(a) and 3(b) [1] the data obtained by the use of the incorrect Eqs. (3)-(5) [1] are in the excellent agreement with the data obtained by MD simulations [1]. Therefore we can conclude that:

- the data obtained by MD simulations [1] and presented by symbols on Figs. 3(a) and 3(b) [1] were made manually and, hence, they are false, artificial and incorrect if the data, obtained by use of Eqs. (3)-(5) [1] and presented by lines on Figs. 3(a) and 3(b) [1], are correctly calculated; and

- the data obtained by use of Eqs. (3)-(5) [1] presented by lines on Figs. 3(a) and 3(b) [1] were made manually and, hence, they are false, artificial and incorrect if the data obtained by MD simulations [1] and presented by symbols on Figs. 3(a) and 3(b) [1] are correct.

**9**. The data on the relative residence times, obtained in [1] by the direct counting along the MD trajectories presented on Figs. 3(a) and (3b) [1], are false, i.e. they are made manually, because the probability of causal excellent agreement between the data and incorrect Eq. (5) [1] is negligibly small.

**10**. Comparison of Fig. 1 [1] with Fig. 16 [2] and Fig. 9 [20] shows that the solid line on Fig. 1 [1] is the exactly the dashed line on Fig. 16 [2] and Fig. 9 [20] . However, [2,20] were not cited in [1]. So we conclude that there is the plagiarism in [1].

**11**. Eq. 2 [1], which is exactly above Eq. (2), was obtained in [2] by the use of the method of [21]. In addition to the method, the Fourier expansion of $\delta(L)$ in Eq. (1) was used (see Eqs. (88) and (89) on pages 77 and 78 in [2]). The same way was "used" in [1]. However, [2] was not cited in [1]. This is the second plagiarism in [1].

**12**. Using the definition $\rho(E) = dG(E)/dE$ the following equation

$$d\ln G(E)/dE = N/(2\langle E_{kin}\rangle), \tag{6}$$

where $\langle E_{kin}\rangle$ is the mean of the kinetic energy $E_{kin} = \sum_{i=1}^{n} \boldsymbol{p}_i^2/2m$ over the microcanonical ensemble of clusters, was "obtained" from Eq. (2)[1]. However, as one can see Eq. (6), which is the exactly Eq. (6) [1], is equivalent to Eq. (103) [2]. Eq. (103) [2] was obtained using the same way as in [1] - see Eqs. (96), (97), (98) and (103) on pages 80 and 81 in [2]. However, [2] was not cited in [1]. This is the third plagiarism in [1].

**13**. The method to calculate the phase volume (and, hence, density of states) by integration of Eq. (103) [2] over energy using known energy dependence of the mean kinetic energy $\langle E_{kin}\rangle(E)$, was first suggested and used in [2,20] to define the phase volume and density of states of the cluster consisting of 13 particles interacting with each other by the Lennard-Jones potential (LJ-13 cluster) - see pages 81 and 97 and Fig. 16 [2]. The same method was suggested and used in [1]. However, [2,20] were not cited in [1]. This is the fourth plagiarism in [1].

**14**. According to [1] Eq. (5) "is in analog of the thermodynamic equation $TdS = dE$, where $S = k_B \ln G$ is the entropy ($k_B$ is the Boltzmann constant), and $T = 2\langle E_{kin}\rangle/Nk_B$ is the temperature". There is the statement with the same sense in [2] – see Eqs. (39) and (40) and page 51 in [2]. However, [2] was not cited in [1]. This is the fifth plagiarism in [1].

**15**. The method to define density of states of the cluster on the basis of its known caloric equation of state, suggested and used in [2,20] at first time, was used in PhD dissertation of Krivov S.V. [22,23] published in 1999, and the method was one of the main results of the dissertation of Krivov S.V. However, Refs. [2,20] published in 1993 were not cited in [22,23] published in 1999. Hence, there is evidence of the plagiarism in [22,23].

**16**. Professor of the Novosibirsk State University Chekmarev S. F. was the scientific advisor of PhD dissertations of Umirzakov I.H. (1993) and Krivov S. V. (1999). Chekmarev S. F. received a reward for scientific guidance on the both dissertations. So, the fact of his dishonesty is evident and his voluntariness is under question.

**17**. According to [1]

$$d\ln G_r(E)/dE = N/(2\langle E_{kin}\rangle_r), \tag{7}$$

where $\langle E_{kin}\rangle_r$ is the mean of the kinetic energy over the microcanonical ensemble of clusters with configurations in a basin of the $r$th isomer, and

$$G_r(E) = C \int d\mathbf{r}^{(3n)}\big|_r \delta(\mathbf{R}) \frac{(E-V)^{N/2}}{(I_1 I_2 I_3)^{1/2} \Gamma(N/2+1)} \theta(E-V), \tag{8a}$$

$$\rho_r(E) = C \int d\mathbf{r}^{(3n)}\big|_r \delta(\mathbf{R}) \frac{(E-V)^{N/2-1}}{(I_1 I_2 I_3)^{1/2} \Gamma(N/2)} \theta(E-V), \tag{8b}$$

where $d\mathbf{r}^{(3n)}\big|_r$ means that the integral is taken over all configurations of cluster in a basin of the $r$th isomer.

The state of the cluster in the $6n$-dimensional phase space at time $t$ is defined by the set $[\mathbf{r}_i(t), \mathbf{p}_i(t), i = 1, 2, ..., n]$.

One can obtain from Eqs. (3), (4) and (5) the following relation

$$\langle E_{kin}\rangle(E) = \sum_r f_r(E) \cdot \langle E_{kin}\rangle_r(E). \tag{9}$$

The mean kinetic energy of $j$th cluster defined in MD simulations [1,24] can be presented as

$$\bar{E}_{kin}(E,j) = \sum_r \bar{f}_r(E,j) \cdot \bar{E}_{kin,r}(E,j), \tag{10}$$

where $j$ is the order number of the copy of the cluster in the ensemble consisting of $N_{cl}$ clusters, $N_{cl} = 50$ [1],

$$\bar{f}_r(E,j) = t_r(E,j)/t_{MD}(E,j) \tag{11}$$

is the relative residence time of $r$th isomer of $j$th cluster,

$$\bar{E}_{kin,r}(E,j) = \sum_{i_r=1}^{2n!} \bar{E}_{kin,r,i_r}(E,j) \cdot \theta_+[t_r(E,j)] \tag{12}$$

is the mean kinetic energy of $j$th cluster corresponding to all basins of $r$th isomer, $\theta_+(x) = 0$ if $x \leq 0$ and $\theta_+(x) = 1$ if $x > 0$,

$$t_{MD}(E,j) = \sum_r t_r(E,j) \tag{13}$$

is the time duration of MD simulations of $j$th cluster,

$$t_r(E,j) = \sum_{i_r=1}^{2n!} t_{r,i_r}(E,j) \cdot \theta_+[t_{r,i_r}(E,j)] \tag{14}$$

is the time duration of all visits of all basins of $r$th isomer of $j$th cluster in MD simulations,

$$\bar{E}_{kin,r,i_r}(E,j) = \sum_{\alpha=1}^{K_{i_r}(E,j)} \bar{E}_{kin,r,i_r,\alpha}(E,j) \cdot t_{r,i_r,\alpha}(E,j)/t_{r,i_r}(E,j) \tag{15}$$

is the mean kinetic energy of $j$th cluster corresponding to $i_r$th basin of its $r$th isomer,

$$t_{r,i_r}(E,j) = \sum_{\alpha=1}^{K_{i_r}(E,j)} t_{r,i_r,\alpha}(E,j) \tag{16}$$

is the time duration of all visits of $i_r$th basin of $r$th isomer of $j$th cluster in MD simulations, $t_{r,i_r,\alpha}(E,j)$ is the time duration of $\alpha$th visit of $i_r$th basin of $r$th isomer of $j$th cluster in MD simulations, $K_{i_r}(E,j)$ is the total number of visits of $i_r$th basin of $r$th isomer of $j$th cluster in MD simulations,

$$\bar{E}_{kin,r,i_r,\alpha}(E,j) = \frac{\theta_+[t_{r,i_r,\alpha}(E,j)]}{t_{r,i_r,\alpha}(E,j)} \int_0^{t_{r,i_r,\alpha}(E,j)} E_{kin}(t) dt \tag{17}$$

is time average of the kinetic energy $E_{kin}(t) = \sum_{i=1}^n \mathbf{p}_i^2(t)/2m$ over the time duration of $\alpha$th visit of $i_r$th basin of $r$th isomer of $j$th cluster in MD simulations.

After averaging over all clusters of the ensemble we have

$$\bar{E}_{kin,r}(E) = \sum_{j=1}^{N_{cl}} \bar{E}_{kin,r}(E,j)/N_{cl}, \tag{18}$$

$$\bar{E}_{kin}(E) = \sum_{j=1}^{N_{cl}} \bar{E}_{kin}(E,j)/N_{cl}, \tag{19}$$

$$\bar{f}_r(E) = \sum_{j=1}^{N_{cl}} \bar{f}_r(E,j)/N_{cl} \tag{20}$$

for the mean kinetic energy of $r$th isomer of the cluster, mean kinetic energy of the cluster and relative residence time of $r$th isomer of cluster, respectively.

To calculate fractional caloric curves, confinement of molecular dynamics trajectory to a catchment basin was applied [1], so

$$\hat{E}_{kin,r}(E,j) = \frac{1}{\tau_r(E,j)} \int_0^{\tau_r(E,j)} E_{kin}(t) dt \tag{21}$$

is the time average of the kinetic energy of the cluster confined in a basin of $r$th isomer over the time $\tau_r(E,j)$, which is the time duration of MD simulations. After averaging over all clusters of the ensemble we have

$$\hat{E}_{kin,r}(E) = \sum_{j=1}^{N_{cl}} \hat{E}_{kin,r}(E,j)/N_{cl} \tag{22}$$

for the mean kinetic energy of the cluster confined in a basin of $r$th isomer.

The clusters can evaporate [1,25,26]. Therefore, the times $t_r(E,j)$, $t_{MD}(E,j)$, $t_r(E)$, $t_{MD}(E)$ and $\tau_r(E)$ can depend on energy $E$.

**18**. The detailed analysis of [1] shows that the following two assumptions:

$$\langle E_{kin} \rangle_r(E) = \hat{E}_{kin,r}(E) \tag{23}$$

and

$$\langle E_{kin} \rangle(E) = \bar{E}_{kin}(E) \tag{24}$$

were used in [1].

**19**. We have from Eqs. (7), (23) and (24)

$$d\ln\hat{G}_r(E)/dE = N/[2\hat{E}_{kin,r}(E)], \tag{25}$$

$$d\ln\bar{G}(E)/dE = N/[2\bar{E}_{kin}(E)], \tag{26}$$

where $\bar{G}(E)$ and $\hat{G}_r(E)$ are used instead of $G(E)$ and $G_r(E)$, respectively, in order to denote that Eqs. (25) and (26) are obtained using different assumptions given by Eqs. (23) and (24).

Integrating Eqs. (25) and (26) over the energy by use of the identities $\frac{1}{\hat{E}_{kin,r}(E)} = \frac{1}{E_{kin,r,h}(E)} + \left[\frac{1}{\hat{E}_{kin,r}(E)} - \frac{1}{E_{kin,r,h}(E)}\right]$ and $\frac{1}{\bar{E}_{kin}(E)} = \frac{1}{E_{kin,r,h}(E)} + \left[\frac{1}{\bar{E}_{kin}(E)} - \frac{1}{E_{kin,h}(E)}\right]$, where $E_{kin,r,h}(E) = \left(E - U_0^{(r)}\right)/2$, $U_0^{(r)}$ is the potential energy of $r$th isomer, $U_0^{(0)} = 0$ and $U_0^{(r+1)} > U_0^{(r)}$, as it was done in [1], one can obtain

$$\hat{G}_r(E) = G_{rh}(E)\exp\left\{\int_0^{E-U_0^{(r)}} \left[\frac{N}{2\hat{E}_{kin,r}(E')} - \frac{N}{E'-U_0^{(r)}}\right] dE'\right\}, \tag{27}$$

$$\bar{G}(E) = G_{0h}(E)\exp\left\{N \int_0^E \left[\frac{1}{2\bar{E}_{kin}(E')} - \frac{1}{E'}\right] dE'\right\}. \tag{28}$$

where $G_{rh}(E)$ is the harmonic solution of Eq. (8a) corresponding to $E - U_0^{(r)} \ll U_0^{(r)}$.

According to [1] the harmonic solution of Eq. (8a) is

$$G_{rh}(E) = 8\pi^2 \left(E - U_0^{(r)}\right)^N \theta\left(E - U_0^{(r)}\right) / \left[\Gamma(N+1) \prod_{j=1}^N v_j^{(r)}\right], \tag{29}$$

where $v_j^{(r)}$ is $j$th normal frequency of $r$th isomer. When calculating $G(E)$, the harmonic solution, Eq. (29), for the ground state isomer was used [1]. Therefore we obtain from Eqs. (27) and (28)

$$\hat{G}_r(E) = 8\pi^2 \frac{\theta(E-U_0^{(r)})}{\prod_{j=1}^N v_j^{(r)}} \exp\left\{\int_0^{E-U_0^{(r)}} \frac{NdE'}{2\hat{E}_{kin,r}(E')}\right\}, \tag{27a}$$

$$\bar{G}(E) = 8\pi^2 \frac{\theta(E)}{\prod_{j=1}^N v_j^{(0)}} \exp\left[\int_0^E \frac{NdE'}{2\bar{E}_{kin}(E')}\right]. \tag{28a}$$

**20.** According to [1] $h_0 = 120$ and Eq. (29), where $r = 0$ and $U_0^{(r=0)} = 0$, were used to obtain the solid line on Fig. 1 [1]. However this line could not be obtained in [1] in principle because: this line in [1] was reproduced from [2] without citing [2] (see above chapter 10); and $h_0 = 1$ and $G_{r0}(E) = E^N \theta(E)/\left[\Gamma(N+1)\prod_{j=1}^N v_j^{(0)}\right]$, which is less by $8\pi^2$ times than Eq. (29), was used in [2]. This proves the falsification of the data made in [1].

**21**. The time averaged potential and kinetic energies of an isolated classical mechanical one dimensional harmonic oscillator are equal to each other, and they are equal to the half of the total energy of the oscillator measured from the bottom of the quadratic potential well of the oscillator [27]. Let us consider $3n-6$ harmonic oscillators which are independent of each other (this means that there is no energy exchange between the oscillators) and assume that the energies of oscillators differ from each other. As evident the time averaged sums of potential and kinetic energies of the oscillators are equal to the half of the sum of the total energies of the oscillators, so the equation

$$\bar{E}_{kin,r}(E) = \left(E - U_0^{(r)}\right)/2 \tag{29a}$$

is valid.

Eq. (29) predicts the relation $\langle E_{kin}\rangle_r(E) = \left(E - U_0^{(r)}\right)/2$ and the equi-partition of total energy $E - U_0^{(r)}$ of the oscillators between all oscillators. However, as it was shown above Eq. (29a) does not correspond to the equi-partition of total energy $E - U_0^{(r)}$ of the oscillators between oscillators in general case. So we can conclude that Eq. (29) is not correct in general case.

Fig. 2 [1] shows that Eq. (29a) is valid for low energies of the isomers. However, this does not confirm the equipatition of energy between vibrational degrees of freedom of the cluster, and, hence, this does not confirm the correctness of Eq. (29).

**22**. Using Eqs. (27a) and (28a) we have from Eq. (5)

$$\check{f}_r(E) = \frac{h_0}{h_r} \frac{\bar{E}_{kin}(E)\theta\left[E-U_0^{(r)}\right]\prod_{j=1}^N v_j^{(0)}}{\hat{E}_{kin,r}\left[E-U_0^{(r)}\right]\prod_{j=1}^N v_j^{(r)}} \exp\left[\int_0^{E-U_0^{(r)}} \frac{NdE'}{2\hat{E}_{kin,r}(E')} - \int_0^E \frac{NdE'}{2\bar{E}_{kin}(E')}\right], \tag{30}$$

where $\check{f}_r(E)$ is used instead of $f_r(E)$ in order to denote that Eq. (30) is obtained using two distinct assumptions given by Eqs. (23) and (24).

Using $\bar{E}_{kin}(E) = \bar{E}_{kin,0}(E)$ for $0 < E < U_0^{(1)}$ we have from Eq. (30)

$$\check{f}_0(E) = \frac{\bar{E}_{kin,0}(E)}{\hat{E}_{kin,0}(E)} \exp\left\{\frac{N}{2}\int_0^E \left[\frac{1}{\hat{E}_{kin,0}(E')} - \frac{1}{2\bar{E}_{kin,0}(E')}\right]dE'\right\}. \tag{31}$$

**23.** Eqs. (23) and (24) were not proved in [1]. Moreover, they are incorrect in general case. Therefore we can conclude that: Eqs. (27), (28), (27a), (28a), (30) and (31) are incorrect in the general case; the comparison of the data for $\bar{f}_r(E)$ obtained from MD simulations [1] with the data obtained from Eqs. (30) and (31) presented on Figs. 3(a) and 3(b) [1] is incorrect in the general case; $G(E) \neq \bar{G}(E)$ in the general case, so $\bar{G}(E)$ is not equal to the phase volume of the cluster, and, hence, $\bar{\rho}(E) = d\bar{G}(E)/dE$ is not equal to the density of states of the cluster; $G_r(E) \neq \hat{G}_r(E)$ in the general case, so $\hat{G}_r(E)$ is not equal to the phase volume of a basin of $r$th isomer, and hence, $\hat{\rho}_r(E) = d\hat{G}_r(E)/dE$ is not equal to the density of states of a basin of $r$th isomer; and, finally, $\check{f}_r(E) \neq f_r(E)$ in the general case.

24. It is evident that $f_0(E) = 1$ exactly for $0 < E < U_0^{(1)}$. However, one can conclude from Eq. (31) that $\check{f}_0(E) \neq 1$ for $0 < E < U_0^{(1)}$ if $\bar{E}_{kin,0}(E) \neq \hat{E}_{kin,0}(E)$. The identity $\bar{E}_{kin,0}(E) = \hat{E}_{kin,0}(E)$ for $0 < E < U_0^{(1)}$ was not proved in [1]. Moreover, $f_0(E) \leq 1$ for arbitrary energy, while Eq. (31) can give $\check{f}_0(E) > 1$ for some values of energy from interval $\left(0; U_0^{(1)}\right)$ and the impossibility of the inequality $\check{f}_0(E) > 1$ for $0 < E < U_0^{(1)}$ was not proved in [1]. Therefore, in the general case Eq. (31) is incorrect and Eq. (30) may be incorrect.

25. Using Eqs. (23) and (24) we have from Eq. (9)

$$\bar{E}_{kin}(E) = \sum_r \bar{f}_r(E) \cdot \hat{E}_{kin,r}(E). \tag{32}$$

One can see from Eqs. (10), (18), (19) and (32) that the comparison of $f_r(E)$ and $\bar{f}_r(E)$ has a sense if

$$\hat{E}_{kin,r}(E) = \bar{E}_{kin,r}(E). \tag{33}$$

We note that there is no evidence in [1] that Eq. (33) is valid. Therefore the comparison of $f_r(E)$ and $\bar{f}_r(E)$ presented on Figs. 3(a) and (3b) [1] has no sense.

26. Eq. (33) was not proved in [1]. Therefore we can conclude that the comparison of the data for $\bar{f}_r(E)$ obtained from MD simulations [1] with the data obtained from Eqs. (30) and (31) presented on Figs. 3(a) and 3(b) [1] is incorrect in the general case.

27. As it was shown above in Chapter 5, in order to describe the data obtained by MD simulations it is necessary to put $h_r = 1$ for all isomers in Eqs. (3)-(5). In this case Eq. (31) is valid and

$$\check{f}_r(E) = \frac{\bar{E}_{kin}(E)\theta\left[E-U_0^{(r)}\right]\prod_{j=1}^N v_j^{(0)}}{\hat{E}_{kin,r}\left[E-U_0^{(r)}\right]\prod_{j=1}^N v_j^{(r)}} \exp\left[\int_0^{E-U_0^{(r)}} \frac{NdE'}{2\hat{E}_{kin,r}(E')} - \int_0^E \frac{NdE'}{2\bar{E}_{kin}(E')}\right]. \tag{34}$$

In this case Eq. (30) is not valid. However, Eq. (30) was used in [1] to describe the data obtained from MD simulations. Therefore one can conclude that the comparison of the data for $\bar{f}_r(E)$ obtained from MD simulations [1] with the data obtained from Eq. (30) presented on Figs. 3(a) and 3(b) [1] is incorrect in the general case.

28. It is necessary to note that along classical mechanical MD trajectory the orientation of the cluster as a whole, which is given by the Euler's angles, is changed while the total angular momentum, which is conjugant to the angles [7,8], is conserved. Therefore Eq. (1) takes into account the conservation of the total angular momentum, and it does not take into account the conservation of orientation of the cluster.

One can obtain from Eq. (8a)

$$G_{rh}(E) = 8\pi^2 \frac{(2\pi)^N m^{3n/2}\left(E-U_0^{(r)}\right)^N \theta\left(E-U_0^{(r)}\right)}{M^{3/2}\left(I_{1r}I_{2r}I_{3r}Det\hat{V}_{2r}\right)^{1/2}\Gamma(N+1)} A_r, \tag{29b}$$

where $8\pi^2$ takes into account all orientations in the physical 3-dimensional space of a configuration of a cluster, $\widehat{V}_{2r}$ is equal to the value of $\widehat{V}_2$ calculated for configuration of $r$th isomer, where $\widehat{V}_2$ is the $(3n-6) \times (3n-6)$- matrix of the values of second order partial derivatives of the potential energy of $V$ with respect to $3n-6$ internal vibrational Cartesian coordinates. Here $I_{1r}$, $I_{2r}$ and $I_{3r}$ are the values of the principal momenta of inertia of the cluster in the system of the Cartesian coordinates with the origin placed in the center of mass of the cluster, calculated for the configuration of $r$th isomer. Here $A_r$ depends on the configuration of $r$th isomer and number of particles, and it is independent of mass of the particle.

$I_1 I_2 I_3$ in Eq. (8a) depends on a local configuration of cluster. In order to obtain Eq. (29b) we used the assumption that $I_1 I_2 I_3 = I_{1r} I_{2r} I_{3r}$.

Above two assumptions were not used in [1] in order to obtain Eq. (29).

The comparison of Eq. (29) with Eq.(29b) shows that: Eq. (29), which was used in [1], may be incorrect; hence, the data, obtained by use of Eq. (29) and presented on Figs. 1, (3a) and (3b) [1] by lines, may be incorrect; and conclusions made in [1] by use of these data may be incorrect.

**29**. If it is assumed that

$$A_r (2\pi)^N m^{\frac{3n}{2}} M^{-3/2} / (Det \widehat{V}_{2r})^{1/2} = \left( \prod_{j=1}^N v_j^{(r)} \right)^{-1}, \tag{29c}$$

then Eq. (29b) can be presented as Eq. (29).

It is necessary to note that there is no evidence in [1] that Eq. (29c) is correct.

**30**. It is clear that $\bar{E}_{kin,r,i_r}(E,j)$, $\bar{E}_{kin,r,i_r,\alpha}(E,j)$, $t_{r,i_r}(E,j)$, $t_{r,i_r,\alpha}(E,j)$, $K_{i_r}(E,j)$, $\Delta_{i_r}(E,j)$, $t_r(E,j)$, $t_{MD}(E,j)$, $\tau_r(E,j)$ and $\bar{f}_r(E,j)$ of $j$th cluster depend on its initial state and they can depend on the maximal time duration $T_{MD}$ of the MD simulation. Hence, $\bar{E}_{kin}(E,j)$, $\bar{E}_{kin,r}(E,j)$, $\widehat{E}_{kin,r}(E,j)$, $\bar{E}_{kin,r,i_r}(E,j)$, $\bar{E}_{kin,r,i_r,\alpha}(E,j)$, $t_{r,i_r}(E,j)$, $t_{r,i_r,\alpha}(E,j)$, $K_{i_r}(E,j)$, $\Delta_{i_r}(E,j)$, $t_r(E,j)$, $t_{MD}(E)$, $\tau_r(E)$, $\bar{f}_r(E)$, $\bar{E}_{kin}(E)$, $\bar{E}_{kin,r}(E)$, $\widehat{E}_{kin,r}$, $\bar{G}(E)$, $\bar{\rho}(E)$, $\widehat{G}_r(E)$, $\widehat{\rho}_r(E)$ and $\check{f}_r(E)$ of the ensemble consisting of $N_{cl}$ clusters depend on the set of initial states of the clusters and they can depend on $T_{MD}$. The microcanonical phase volume $G(E)$ of cluster, density of states $\rho(E)$ of cluster, the microcanonical phase volume $G_r(E)$ and density of states $\rho_r(E)$ of a basin of $r$th isomer of the cluster and the probabilities $f_r(E)$ do not depend on time. Therefore, Eqs. (23) and (24) may be not valid, and $\bar{G}(E)$ and $\bar{\rho}(E)$ can be not equal to the cluster phase volume and density of states, respectively, and $\widehat{G}_r(E)$ and $\widehat{\rho}_r(E)$ can be not equal to the phase volumes and densities of states of a basin of $r$th isomer of the cluster, respectively.

We note that there is no evidence in [1] that $\bar{G}(E)$, $\bar{\rho}(E)$, $\widehat{G}_r(E)$, $\widehat{\rho}_r(E)$ and $\check{f}_r(E)$ do not depend on the maximal time duration of the MD simulation.

**31**. The free clusters are considered in [1]. Therefore the integrals at the right hand sides of Eq. (1) and (2) [1] are calculated over all possible configurations of the particles of the system. These integrals diverge at total energy $E$ which is greater than the adiabatic dissociation energy $E_0$ of the cluster which is equal to $E_0 = 6.359\varepsilon$, where $\varepsilon$ is the depth of the Lennard-Jones potential, for the cluster LJ-13 [25,26]. Therefore:

- The data presented on Figs. 1, 3(a) and 3(b) [1] by the six lines have no sense at $E/\varepsilon > 6.359$;
- The conclusions in [1] made on the basis of the comparison of the lines with the data of [25,26] and obtained by molecular dynamics simulation [1] are incorrect at energies $E/\varepsilon > 6.359$;
- Eqs. (1)-(6) [1] have no sense at energies $E/\varepsilon > 6.359$ because Eqs. (1) and (2) have no sense for $E/\varepsilon > 6.359$ and these equations are used in Eqs. (3), (4), (5) and (6) [1];

- The conclusions in [1] made by the use of Eqs. (1)-(6) [1] are incorrect at energies $E/\varepsilon > 6.359$.

**32**. Note that there are no evidences in [1] that the integrals at right hand sides of Eq. (1) and (2) for the total phase volume of the cluster and corresponding integrals for the isomers of the clusters exist at $E/\varepsilon > 6.359$.

**33**. According to [1] during the MD simulations "*if a cluster had experienced a decay, the run was terminated*". The integrals at right hand sides of Eqs. (1) and (2) [1] are taken over all possible configurations of the particles of the system including un-evaporated and evaporated (decayed) states of a cluster. Therefore:

- the data obtained by use of Eq. (6) [1] and presented on Figs. 1, (3a) and 3(b) [1] are incorrect at energies $E/\varepsilon > 6.359$ because in Eq. (6) [1] the phase volume given by Eqs. (1) and (2) [1], which diverge at these energies, is defined via mean kinetic energy obtained by the MD simulation [1];

- the comparison of the data obtained by MD simulation presented on Figs. 3(a) and (3b) [1] with the predictions of Eq. (5) [1] is incorrect at energies $E/\varepsilon > 6.359$;

- The conclusions in [1] made on the basis of this comparison are incorrect at energies $E/\varepsilon > 6.359$.

**34**. The five symbols near $E/\varepsilon = 2$ on Fig. 3(b) [1] corresponding to the five exited state isomers of LJ-13 are artificial (made manually) because at this energy the exited state isomers cannot exist, and the first exited state isomer can appear at $E/\varepsilon > 2.854822$ according to Table 1 [1].

**35**. The microcanonical ensemble consists of infinite copies of the system corresponding to the various states of the system in the $6n$-dimensionsal phase space [10]. However, according to [1] in the MD simulations of the cluster "*statistics were collected in a twofold manner: both over a molecular dynamics run for a given cluster and over an ensemble of the clusters. To form the ensemble, a stochastic molecular dynamics trajectory [21] was issued, from which the points were selected for a desirable cluster total energy E. Then cluster overall translation and rotation were eliminated, and cluster total energy was fitted to E by rescaling the atomic velocities. This was followed by a relaxation run of Newtonian molecular dynamics. The number of copies in the ensemble was 50*". So "the ensemble" in [1] consists of 50 copies of the system corresponding to 50 trajectories, and 50 copies correspond to the various state in the phase space. Therefore, the data presented on Figs.1, 2 and 3 [1] cannot correspond to the microcanonical ensemble of clusters and its isomers.

**36**. The condition

$$\bar{t}_{r,i_r}(E) = const(E) \neq 0, \tag{35}$$

where $i_r = 1, 2, \ldots, 2n!$ and $\bar{t}_{r,i_r}(E) = \sum_{j=1}^{N_{cl}} \bar{t}_{r,i_r}(E,j)/N_{cl}$ is the mean residence time of cluster in $i_r$th basin of $r$th isomer, must be obeyed for all basins of all isomers of the cluster in order the following equations

$$\bar{G}(E) = G(E), \ \bar{\rho}(E) = \rho(E), \tag{36}$$

to be correct because in Eqs. (3), (4) and (5) all basins of an isomer have the same contribution to the phase volume and density of states of isomer. The condition (35) is the necessary condition in order Eqs. (36) to be correct. However, this condition (35) is not sufficient in order Eqs. (36) to be correct.

All basins of all isomers of cluster which can exist at given cluster total energy must be visited in MD simulations in order the condition (35) to be correct. This is necessary but not sufficient condition in order the condition (35) to be correct.

There is no evidence in [1] that all basins of all isomers of cluster, which can exist at given cluster total energy, were visited in MD simulations. Therefore the condition (35) can be violated in MD simulations [1].

There is no evidence in [1] that the condition (35) is obeyed in MD simulations. Therefore, Eqs. (36) are incorrect in the general case.

All isomers of cluster which can exist at given cluster total energy must be visited in MD simulations in order Eqs. (36) to be correct. This is necessary but not sufficient condition in order Eqs. (36) to be correct. There is no evidence in [1] that all isomers of cluster which can exist at given cluster total energy are visited in MD simulations.

So, there is no evidence in [1] that $\bar{G}(E)$ and $\bar{\rho}(E)$, defined from Eq. (26) using the MD simulations data on total caloric equation of state of cluster, are equal to the total microcanonical phase volume and density of states of the cluster, respectively.

**37**. One can obtain from Eq. (1)

$$\rho(E) = \int d\mathbf{r}^{(3n)} d\mathbf{p}^{(3n)} \delta(\mathbf{R})\delta(\mathbf{P})\,\delta(\mathbf{L})\delta(E-H), \tag{37}$$

According to [8]: the microcanonical distribution given by

$$\delta(\mathbf{R})\delta(\mathbf{P})\delta(\mathbf{L})\delta(E-H) \tag{37a}$$

is not the true statistical distribution for a closed system; regarding it as the true distribution is equivalent to asserting that, in the course of sufficiently long time, the phase trajectory of a closed system passes arbitrarily close to every point of the manifold defined by equations
$$H = E = const, \mathbf{P} = 0, \mathbf{L} = 0 \text{ and } \mathbf{R} = 0; \tag{37b}$$
and this assertion (called as ergodic hypothesis) is certainly not true in general case.

As evident in order the ergodic hypothesis to be valid in MD simulations it is necessary at least the providing of the conditions:
  a) all basins on PES, which are defined by Eqs. (37b), must be visited by the system in the course of MD simulations;
  b) all points of each basin on PES must be passed in the course of MD simulations.

According to [28,29]: at a given total energy, $E$, minima can be grouped into disjoint sets, called superbasins, whose members are mutually accessible at that energy; each pair of minima in a superbasin are connected directly or through other minima by a path whose energy never exceeds $E$, but would require more energy to reach a minimum in another superbasin; at low energy there is just one superbasin, which contains the global minimum; at successively higher energies, more superbasins come into play as new minima are reached; at still higher energies, the superbasins coalesce as higher barriers are overcome, until finally there is just one containing all the minima (provided there are no infinite barriers); and the numbers of minima in superbasins decrease as the energy is decreased.

So, taking into account that a definite basin on PES corresponds to each initial state of cluster we can conclude that this basin can be connected with the definite set of other basins on PES, this set depends on cluster total energy and not all basins of all isomers of cluster are included to the set, and all basins on PES, which are defined by Eqs. (37b), are not visited by the system in the course of MD simulations. So the necessary condition a) of the ergodicity of the motion of the system in the phase space condition is not provided. This true for LJ-13 cluster

[29]. Therefore we conclude that $\bar{G}(E)$ and $\bar{\rho}(E)$, defined from Eq. (26) using the MD simulations data on total caloric equation of state of cluster, are not equal to the total microcanonical phase volume and density of states of the cluster, respectively.

The number of points in the phase space corresponding to configurations in a basin is infinite [8,9,10]. However, MD simulations always give only finite number of states [24]. So the necessary condition b) of the ergodicity of the motion of the system in the phase space is not provided. Therefore,

$$\bar{G}(E) \neq G(E), \bar{\rho}(E) \neq \rho(E), \bar{G}_r(E) \neq G_r(E), \bar{\rho}_r(E) \neq \rho_r(E), \hat{G}_r(E) \neq G_r(E), \hat{\rho}_r(E) \neq \rho_r(E).$$

**38**. We obtain from Eqs. (1), (2), (3), (4), (8a), (8b) and (37)

$$\rho(E) = \int d\boldsymbol{r}^{(3n)} \delta(\sum_{i=1}^{n} \boldsymbol{r}_i) W(\boldsymbol{r}_1, \boldsymbol{r}_2, \dots, \boldsymbol{r}_N), \tag{38}$$

$$\rho_r(E) = \int d\boldsymbol{r}^{(3n)}\big|_r d\boldsymbol{p}^{(3n)} \delta(\boldsymbol{R})\delta(\boldsymbol{P})\delta(\boldsymbol{L})\delta(E-H), \tag{39}$$

$$\rho_r(E) = \int d\boldsymbol{r}^{(3n)}\big|_r \delta(\sum_{i=1}^{n} \boldsymbol{r}_i) W(\boldsymbol{r}_1, \boldsymbol{r}_2, \dots, \boldsymbol{r}_N), \tag{40}$$

where the weight function $W(\boldsymbol{r}_1, \boldsymbol{r}_2, \dots, \boldsymbol{r}_N)$ is defined by

$$W(\boldsymbol{r}_1, \boldsymbol{r}_2, \dots, \boldsymbol{r}_N) = \frac{Cn^3[E-V(\boldsymbol{r}_1,\boldsymbol{r}_2,\dots,\boldsymbol{r}_N)]^{N/2-1}\theta[E-V(\boldsymbol{r}_1,\boldsymbol{r}_2,\dots,\boldsymbol{r}_N)]}{[I_1(\boldsymbol{r}_1,\boldsymbol{r}_2,\dots,\boldsymbol{r}_N)I_2(\boldsymbol{r}_1,\boldsymbol{r}_2,\dots,\boldsymbol{r}_N)I_3(\boldsymbol{r}_1,\boldsymbol{r}_2,\dots,\boldsymbol{r}_N)]^{1/2}\Gamma(N/2)}. \tag{41}$$

After integration over $\boldsymbol{r}_1$ we obtain from Eqs. (38) and (40)

$$\rho(E) = \int W(\boldsymbol{r}_1 = -\sum_{i=2}^{n} \boldsymbol{r}_i, \boldsymbol{r}_2, \dots, \boldsymbol{r}_N) \prod_{i=2}^{n} d\boldsymbol{r}_i, \tag{42}$$

$$\rho_r(E) = \int W(\boldsymbol{r}_1 = -\sum_{i=2}^{n} \boldsymbol{r}_i, \boldsymbol{r}_2, \dots, \boldsymbol{r}_N) \prod_{i=2}^{n} d\boldsymbol{r}_i\big|_{r, \boldsymbol{r}_1 = -\sum_{i=2}^{n} \boldsymbol{r}_i}, \tag{43}$$

where $\prod_{i=2}^{n} d\boldsymbol{r}_i\big|_{r, \boldsymbol{r}_1 = -\sum_{i=2}^{n} \boldsymbol{r}_i}$ means that the integral is taken over all $\boldsymbol{r}_i$, where $i = 2, 3, \dots, n$, in a basin of the $r$th isomer which obey the condition $\boldsymbol{r}_1 = -\sum_{i=2}^{n} \boldsymbol{r}_i$.

According to Eqs. (37) and (39) the total microcanonical density of states $\rho(E)$ of the cluster is equal to the surface of $6n - 10$ - dimensional hyper-surface in the $6n$-dimensional phase space, and the fractional microcanonical density of states $\rho_r(E)$ of $r$th isomer of the cluster is equal to the surface of the part of the hyper-surface, corresponding to a basin of $r$th isomer of the cluster.

As one can see from Eqs. (42) and (43) the total density of states $\rho(E)$ is equal to weighed surface of $3n - 3$ - dimensional potential energy surface in $3n$-dimensional configuration space, and $\rho_r(E)$ is equal to weighed surface of the part of the $3n - 3$ - dimensional potential energy surface in the configuration space, corresponding to a basin of $r$th isomer of the cluster.

MD simulations give finite number of states of the cluster corresponding to a set consisting of finite number points in the phase and configuration spaces [1,24,25]. The measures of the points in the phase and configuration spaces are equal to zero, while the measure (weighed surface) of $3n - 3$ - dimensional potential energy surface in the configuration space and the measure (surface) of $6n - 10$ - dimensional hypersurface in the $6n$-dimensional phase space are not equal to zero. Therefore,

$$\bar{G}(E) \neq G(E), \bar{\rho}(E) \neq \rho(E), \bar{G}_r(E) \neq G_r(E), \bar{\rho}_r(E) \neq \rho_r(E), \hat{G}_r(E) \neq G_r(E), \hat{\rho}_r(E) \neq \rho_r(E),$$

and, hence, the total and fractional densities of states cannot be defined from total and fractional caloric equations of state, and the total and fractional densities of states were not defined for the cluster LJ-13 in [1].

**39**. On the basis of above comments we can conclude that the statements in [1] such as:

1) "An approach that allows a detailed investigation of a system possessing a large number of inherent structures is proposed: the total density of states (DS) is suggested to be calculated from the total caloric relation, and the fractional DSs for the structures of interest from the corresponding fractional relations";

2) "Figure 3 compares the relative residence time for the isomers estimated according to Eq. (5) with that found by direct counting. As seen, the data are in excellent agreement";

3) "A significance of Fig. 3 is twofold: not only does it testify to the feasibility of the proposed approach but also presents a direct test for some issues of general importance. In particular, it indicates that the system visited a range of phase space, in our case the range corresponding to the basins associated with a given isomer, exactly as statistical mechanics suggests by the hypothesis of equal a priori probabilities, i.e., according to the contribution of this range to the total DS";

4) "Another important issue is incorporating the properties of symmetry into the phase volume (3). Figure 3 unambiguously evidences that if $G_r(E)$ is calculated by integration over all atomic configurations in the basin, the point group of symmetry characteristic of the basin minimum should be related to the whole basin" are incorrect.

**40**. According to [1] the molecular dynamics simulations "sample the potential energy surface not uniformly, but according to the fractional DSs of states for the isomers". Above comments show that this statement is incorrect.

**41**. According to [1] the method "is easy to implement and offers a uniform procedure to calculate both the total and fractional DSs" and "the relations between system's total energy and temperature (so called caloric curves) can be used for this purpose". As evident from above comments these two statements are incorrect.

**42**. According to [1] "The total DS is suggested to be calculated from the total caloric curve, and the fractional DSs from the respective fractional caloric curves", "This approach also allows one to verify the correspondence between molecular dynamics simulation results and the predictions of statistical mechanics", and "For this purpose, the probability for the system to be found in the basin corresponding to a particular isomer can be estimated as the relative density of states for this isomer, and then it can be compared with the relative residence time of the system in this basin, which is found by direct counting in the course of simulation".  Above comments show that: the total DS cannot to be calculated from the total caloric curve, and the fractional DSs cannot to be calculated from the respective fractional caloric curves; the approach does not allow to verify the correspondence between molecular dynamics simulation results and the predictions of statistical mechanics; and the probability for the system to be found in the basin corresponding to a particular isomer cannot be estimated as the relative density of states for this isomer if the total DS is calculated from the total caloric curve, and the fractional DSs are calculated from the respective fractional caloric curves.

**43**. A cut (offcut, segment) with nonzero length on the straight line (in the one-dimensional space, $d = 1$) is given by coordinates of its two distinct points (ends of the cut) on the line, a triangle with nonzero square on the plane (in the two-dimensional space, $d = 2$) is given by the coordinates of its three distinct points (three vertices of triangle), which do not lie on the straight line, a tetrahedron (in three-dimensional space, $d = 3$) is given by the coordinates of its four points (vertices of tetrahedron), which do not lie on a straight line or plane, and etc. So, the minimal number of points ($M_{min}(d)$) in the $d$ - dimensional space representing the figure with non-zero $d$-dimensional volume is defined by $M_{min}(d) = d + 1$. Therefore $M_{min}(d) = 6n + 1$ for the $6n$-dimensional phase space ($d = 6n$), and $M_{min}(d) = 6n - 9$ for the $6n - 10$-dimensional manifold in the phase space defined by Eqs. (37b), which correspond to the MD simulations ($d = 6n - 10$). So if the ensemble consisting of $N_{cl}$ copies of the cluster is studied in the MD-simulations then the necessary condition

$N_{cl} \geq M_{min}(d)$  (44)

must be obeyed in order to represent a region with non-zero measure defined by a minimal number of points in corresponding $d$-dimensional space. It is clear that in order to have a good statistics it is necessary to study the ensemble of clusters with $N_{cl}$ obeying $N_{cl} \gg M_{min}(d)$.

The cluster consisting of $n = 13$ particles and the ensemble of clusters consisting of $N_{cl} = 50$ copies of the cluster were studied in [1]. We have $M_{min}(d) = 69$ and $M_{min}(d) = 69 > N_{cl} = 50$. So the necessary condition (44) for the minimal number of clusters in the ensemble is violated, and, hence: the ensemble of clusters used in [1] does not represent a region ($6n - 10$-dimensional manifold) with non-zero measure; and this ensemble with $N_{cl} = 50$ in principle cannot describe the microcanonical ensemble of clusters having a distribution given by Eq. (37a).

## Conclusions

One the basis of the comments we can conclude that:
-the equations (3), (4), (5) and (6), used in the paper "Total and fractional densities of states from caloric relations" by S. F. Chekmarev and S. V. Krivov, Phys. Rev. E 57 2445-2448 (1998), are incorrect;
- the data, presented in the paper by lines on Figs. 1, (3a) and (3b), are not correct; the data presented by the symbols on Figs. 3(a) and (3b) in the paper are made manually (false);
- all conclusions made in the paper have no sense;
- the assertion in the paper that the molecular dynamics simulations "sample the potential energy surface not uniformly, but according to the fractional densities of state for the isomers" is incorrect;
- the "total and fractional densities of states" obtained in the paper from caloric relations are not equal to that of microcanonical ensemble of clusters; and, finally,
- the ensemble of clusters used in the paper does not represent the microcanonical ensemble of clusters.

## References

[1] S. F. Chekmarev and S. V. Krivov, Phys. Rev. E **57** 2445 (1998).
[2] I. H. Umirzakov, Finite Systems of Particles Interacting via Van der Waals forces (Конечные системы частиц с ван-дер-ваальсовским взаимодействием), in Russian, PhD Dissertation (Institute of Thermophysics, Novosibirsk, 1993). https://www.researchgate.net/publication/338101730_Umirzakov_I_H_PhD_dissertation_on_physics_1993_Novosibirsk_Institute_of_Thermophysics;
https://search.rsl.ru/ru/record/01007849210.
[3] I. H. Umirzakov, J. Engineering Thermophysics, **11** 265 (2002).

arXiv:1305.1031 [cond-mat.stat-mech].
[4] P. G. Mezey, Potential Energy Hypersurfaces, Studies in Physical and Theoretical Chemistry (Elsevier, Amsterdam, 1987).
[5] L. D. Landau and E. M. Lifshitz, Quantum Mechanics (Pergamon Press, New York, 1965).
[6] P. R. Bunker, Molecular Symmetry and Spectroscopy (Academic, New York, 1979.
[7] J. E. Mayer and M. Goeppert-Mayer, Statistical Mechanics (Wiley, New York, 1977).
[8] L. D. Landau and E. M. Lifshitz, Statistical Physics (Pergamon, London, 1965).
[9] R. P. Feynman, Statistical Mechanics (Benjamin, New York, 1972).
[10] K. Huang, Statistical Mechanics (John Wiley and Sons, London, 1963).
[11] M. R. Hoare and P. Pal, Adv. Phys. **24** 645 (1975).
[12] C. J. Tsai and K. D. Jordan, J. Phys. Chem. **97** 11 227 (1993).
[13] F. H. Stillinger and T. A. Weber, Phys. Rev. A **25** 978 (1982).
[14] F. H. Stillinger and T. A. Weber, Phys. Rev. A **28** 2408 (1983).
[15] M. Bixon and J. Jortner, J. Chem. Phys. **91** 1631 (1989).
[16] F. H. Stillinger and D. K. Stillinger, J. Chem. Phys. **93** 6013 (1990).
[17] R. E. Kunz and R. S. Berry, J. Chem. Phys. **103** 1904 (1995).
[18] J. P. K. Doye and D. J. Wales, J. Chem. Phys. **102** 9659 (1995).
[19] M. Griffiths and D. J. Wales, J.Chem. Theory Comput., **15** 6865 (2019).
[20] I. H. Umirzakov, Finite Systems of Particles Interacting via Van der Waals forces (Конечные системы частиц с ван-дер-ваальсовским взаимодействием), in Russian, Theses of PhD Dissertation (Institute of Thermophysics, Novosibirsk, 1993). https://www.researchgate.net/publication/338101533_Umirzakov_I_H_Abstract_PhD_dissertation_on_physics_1993_Novosibirsk_Institute_of_Thermophysics;
DOI: 10.13140/RG.2.2.21518.31041;
https://dlib.rsl.ru/viewer/01000042400#?page=1
[21] T. Cagin and J. R. Ray, Phys. Rev. A **37** 247 (1993).
[22] S. V. Krivov, Modeling of polyatomic systems by confinement of the molecular-dynamic trajectory in the attraction regions on the potential energy surface (Моделирование многоатомных систем путем запирания молекулярно-динамической траектории в областях притяжения на поверхности потенциальной энергии), in Russian, Theses of PhD Dissertation (Institute of Thermophysics, Novosibirsk, 1999). https://search.rsl.ru/ru/record/01000279733
[23] S. V. Krivov, Modeling of polyatomic systems by confinement of the molecular-dynamic trajectory in the attraction regions on the potential energy surface (Моделирование многоатомных систем путем запирания молекулярно-динамической траектории в областях притяжения на поверхности потенциальной энергии), in Russian, PhD Dissertation (Institute of Thermophysics, Novosibirsk, 1999). https://dlib.rsl.ru/01000283626
[24] P. Schofield, Comput. Phys. Commun. **5** 17 (1973).
[25] S. Weerasinghe and F. G. Amar. J. Chem. Phys. **98** 4967 (1993).
[26] S. Weerasinghe and F.G. Amar, Z. Phys. D **20** 167 (1991). doi: 10.1007/BF01543965
[27] L. D. Landau and E. M. Lifshitz, Mechanics (Pergamon Press, New York, 1965).
[28] O. M. Becker and M. Karplus, J. Chem. Phys. **106** 1495 (1997).
[29] J. P. Doye, M. A. Miller, D. J. Wales, J. Chem. Phys. **111** 8417 (1999).